\PassOptionsToPackage{unicode}{hyperref}
\PassOptionsToPackage{hyphens}{url}

\documentclass{article}
\usepackage[final,nonatbib]{neurips_2024}

\makeatletter
\renewcommand{\@noticestring}{}
\makeatother

\usepackage[utf8]{inputenc}
\usepackage[T1]{fontenc}
\usepackage{hyperref}
\usepackage{url}
\usepackage{booktabs}
\usepackage{amsfonts}
\usepackage{nicefrac}
\usepackage{microtype}

\usepackage{xcolor,colortbl}
\usepackage{amsmath,amssymb,amsthm}
\usepackage{graphicx}
\usepackage{wrapfig}
\usepackage{subcaption}
\usepackage{enumitem}
\usepackage{multirow}
\usepackage{threeparttable}
\usepackage{xspace}
\usepackage{tikz}
\usepackage{pgfplots}
\pgfplotsset{compat=1.18}
\usepackage{float}

\usepackage{longtable,array}
\usepackage{calc}       
\usepackage{etoolbox}   
\graphicspath{{assets/}{assets/media/}}

\makeatletter
\newsavebox\pandoc@box
\newcommand*\pandocbounded[1]{%
  \sbox\pandoc@box{#1}%
  \Gscale@div\@tempa{\textheight}{\dimexpr\ht\pandoc@box+\dp\pandoc@box\relax}%
  \Gscale@div\@tempb{\linewidth}{\wd\pandoc@box}%
  \ifdim\@tempb\p@<\@tempa\p@\let\@tempa\@tempb\fi
  \ifdim\@tempa\p@<\p@\scalebox{\@tempa}{\usebox\pandoc@box}%
  \else\usebox\pandoc@box%
  \fi}
\makeatother

\usepackage{amsmath,amsfonts,bm}

\def\eqref#1{equation~\ref{#1}}

\def\1{\bm{1}}

\DeclareMathAlphabet{\mathsfit}{\encodingdefault}{\sfdefault}{m}{sl}
\SetMathAlphabet{\mathsfit}{bold}{\encodingdefault}{\sfdefault}{bx}{n}

\title{Real-Time RAG for the Identification\\ of Supply Chain Vulnerabilities}

\author{%
  Jesse Ponnock \\
  MITRE Corporation \\
  \texttt{jponnock@mitre.org}
  \And
  Grace Kenneally \\
  MITRE Corporation \\
  \texttt{gkenneally@mitre.org}
  \And
  Michael Robert Briggs \\
  MITRE Corporation \\
  \texttt{mbriggs@mitre.org}
  \And
  Elinor Yeo \\
  MITRE Corporation \\
  \texttt{eyeo@mitre.org}
  \And
  Tyrone Patterson III \\
  MITRE Corporation \\
  \texttt{tpatterson@mitre.org}
  \And
  Nicholas Kinberg \\
  MITRE Corporation \\
  \texttt{nkinberg@mitre.org}
  \And
  Matthew Kalinowski \\
  MITRE Corporation \\
  \texttt{mkalinowski@mitre.org}
  \And
  David Hechtman \\
  MITRE Corporation \\
  \texttt{dhechtma@mitre.org}
}

\begin{document}
\maketitle

\begin{center}
\vspace*{-1.5\baselineskip}
\small\emph{Originally prepared March 2025; submitted to arXiv August 2025.}
\vspace*{1.0\baselineskip}
\end{center}

\begin{abstract}
New technologies in generative AI can enable deeper analysis into our
nation's supply chains but truly informative insights require the
continual updating and aggregation of massive data in a timely manner.
Large Language Models (LLMs) offer unprecedented analytical
opportunities however, their knowledge base is constrained to the
models' last training date, rendering these capabilities unusable for
organizations whose mission impacts rely on emerging and timely
information. This research proposes an innovative approach to supply
chain analysis by integrating emerging Retrieval-Augmented Generation
(RAG) preprocessing and retrieval techniques with advanced web-scraping
technologies. Our method aims to reduce latency in incorporating new
information into an augmented-LLM, enabling timely analysis of supply
chain disruptors. Through experimentation, this study evaluates the
combinatorial effects of these techniques towards timeliness and quality
trade-offs. Our results suggest that in applying RAG systems to supply
chain analysis, fine-tuning the embedding retrieval model consistently
provides the most significant performance gains, underscoring the
critical importance of retrieval quality. Adaptive iterative retrieval,
which dynamically adjusts retrieval depth based on context, further
enhances performance, especially on complex supply chain queries.
Conversely, fine-tuning the LLM yields limited improvements and higher
resource costs, while techniques such as downward query abstraction
significantly outperforms upward abstraction in practice.
\end{abstract}

\section{Background}

Recently, the U.S. government has included a call to study supply chain
vulnerability in its efforts to understand overall U.S. economic and
national security. In the presidential memorandum regarding
\emph{America First Trade Policy} {[}1{]}, ``stable supply chains'' are
noted as a component of national security. The memo also requires the
federal government to understand how supply chains might be used to
circumvent U.S. policy towards countries that engage in unfair trade
practices. In a separate presidential memorandum on \emph{Unleashing
American Energy} {[}2{]}, one reason given for the need to encourage
U.S. non-fuel mineral production is to ``strengthen supply chains for
the United States and its allies and reduce the global influence of
malign and adversarial states.'' Additionally, new U.S. laws, such as
the Uyghur Forced Labor Prevention Act, require U.S. importers to know
their full supply chains, down to raw material extraction/production
{[}3{]}.

However, granular and timely supply chain data is difficult to obtain.
Economic literature often uses national-level trade or input-output
statistics to analyze global supply chains {[}4{]}. This data grossly
aggregates commodities and businesses which could mask important
information about key products. The data also is usually published with
a considerable time delay; for example, as of February 2025, the most
recent year for the Global Trade Analysis Project's global input-output
dataset is 2017. While this dataset has over 140 counties, it aggregates
data to 65 industry sectors {[}5, 6{]}. It is possible to use more
detailed customs data, like those from ImportGenius, that contains
firm-level trade transactions. However, this is usually obtained from
expensive data providers and it is prohibitively time-intensive to
combine with data from other countries {[}7{]}.

There are less used, though potentially rich source of publicly
available supply chain data that have tremendous value but are
challenging to scrape for pertinent information, such as mandatory
regulatory filings. The Securities Exchange Commission (SEC) requires
certain public disclosures for any firm listed on a stock exchange.
These public filings contain key data, such as risks to operations, key
suppliers and customers, and beneficial ownership. Until recently,
aggregating the data was only feasible by specialized firms that would
then re-sell the data, most notably Bloomberg. But even with these data
aggregators, extracting supply chain information remains a manual task
because the information is usually contained in narrative form instead
of in structured data tables.

The advent of LLMs has opened new possibilities for supply chain data
collection. Since LLMs can ingest large amounts of text-based data, they
can potentially plug in the holes in supply chain data, especially if
they are structured to incorporate recent information. These holes
include the lag between data publication and analysis and of missing the
proverbial ``needle in the haystack'' in long public statements. LLMs
can solve each of these problems with immediate, granular, analysis,
effectively identifying critical insights that human analysts might
miss. This study explores the possibility that by applying various
augmentation techniques, an LLM can answer questions of critical
importance to policymakers regarding U.S. supply chains, by searching
for answers across firms' recent SEC filings.

\section{Problem Statement}

\subsection{Problem Statement}

In recent years, LLMs have gained increasing attention for their ability
to reason over vast amounts of text-based information while RAG has
excelled at augmenting the parametric memory of these models with
external knowledge bases {[}8{]}. The role of the LLM in this context is
to maximize the probability of predicting the correct end to a sequence,
often in response to a prompt. This is generally viewed as an
optimization problem provided through the formalization below:

\begin{equation}
p_{\theta}\bigl(y_i \mid x, y_{1:i-1}\bigr)
\label{eq:next-token}
\end{equation}

Here we aim to maximize the probability of a predicted token \(y_{i}\),
where a token is a word or a part of a word that is part of a prompt.
The subscript \(i\) denotes the position of the token in the sequence.
Given a prompt \emph{x} and all previous tokens \(y_{1:i\  - \ 1}\), the
model conditions on the prompt and prior context to capture the most
likely sequence of tokens, ensuring that each new prediction remains
consistent with the given context and the model's learned distribution.
RAG extends this concept further by integrating retrieval which forms a
new joint probability as follows.

\begin{equation}
p_{\eta}(z \mid x)\, p_{\theta}\bigl(y_i \mid x, z, y_{1:i-1}\bigr)
\label{eq:rag}
\end{equation}

Similar to LLMs, we also typically view RAG as a maximization problem
though here we are attempting to locate an optimum joint probability
between the likelihood, \(p_{\eta}\), of locating the correct supporting
information, \emph{z}, and predicting the correct token sequence
\emph{y} conditioned on both the prompt, \emph{x,} and the correct
retrieved information, \emph{z}. With these foundations established, we
formalize our problem as follows.

\begin{equation}
T_{\text{event}}
-\min_{\tau}\Bigl\{\tau:\,
\mathcal{M}\!\bigl(\tau;\, p_{\eta}(z \mid x),\, p_{\theta}(y_i \mid x, z, y_{1:i-1})\bigr)
\ge \mathcal{M}_{\text{target}}\Bigr\}
\label{eq:objective}
\end{equation}

This work aims to identify a minimum time \(\tau\) such that the
distance is minimized between an event happening in the world at time
\(T_{\text{event}}\) and the ability for an LLM to generate a response
\(y_{i}\) using supporting information \(z\), equal to or greater
than a quality threshold, represented here as \(\mathcal{M}_{\text{target}}\). By
minimizing the distance between \(T_{\text{event}}\) and \(\tau\) we can get
closer to achieving a real-time system, capable of reasoning against
information and events without the knowledge gap inherent to LLMs and
traditional RAG implementations. Further, by instantiating a quality
threshold we ensure that speed is not achieved at the cost of usability
and that the techniques deployed maintain a threshold for effectiveness
in a task. This, in turn, provides a formalization for trade space
analysis in advanced LLM augmentation that balances both timeliness and
quality.

\section{Methods}

The crux of the optimization efforts in our RAG pipeline lies in the
usage of various RAG techniques meant to improve effectiveness over a
standard RAG implementation. Our review of the literature around these
methods has shown that they mostly fall into five categories:
pre-retrieval, at-retrieval, post-retrieval, generation, and
augmentation.

\begin{figure}[htbp]
\centering
\includegraphics[width=\linewidth]{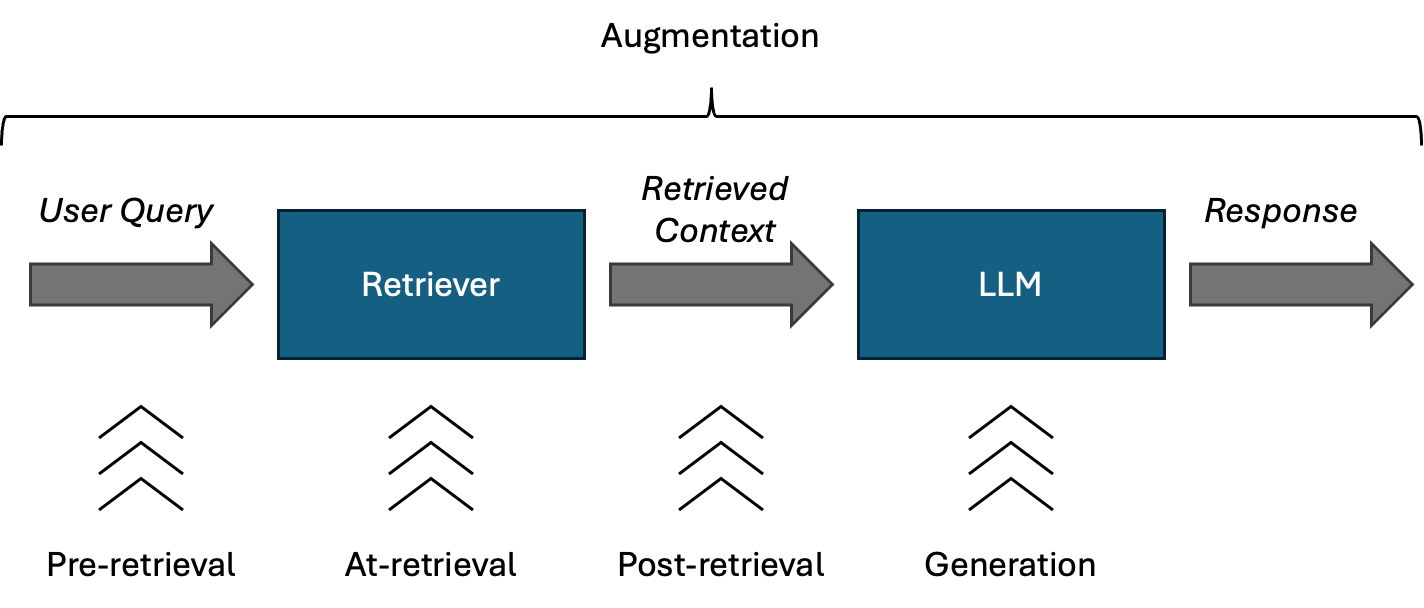}
\caption{RAG Optimization Overview}
\end{figure}

Pre-retrieval techniques focus on enhancing the quality of the retrieval
process before querying the vector store. These methods often involve
improving query formulation through query expansion, rephrasing, or
embedding refinement to better capture the intent and context of the
query. Post-retrieval techniques, on the other hand, aim to refine the
retrieved documents after the initial retrieval step. This can include
re-ranking passages based on relevance scores, filtering out low-quality
results, or leveraging cross-encoder models for more precise ranking.
At-retrieval techniques focus on optimizing the retrieval process
itself, including fine-tuning the embedding model to better align query
and document representations, as well as exploring different retrieval
paradigms. Generation techniques concentrate on improving the LLM's
response generation by incorporating retrieved knowledge more
effectively and enhancing the synthesis of retrieved information into
coherent outputs. Finally, augmentation techniques encompass approaches
that extend or modify the standard RAG architecture to improve system
performance or expand functionality. These methods often involve
integrating alternative retrieval structures, such as graph-based
retrieval in GraphRAG {[}31{]}, or incorporating multimodal inputs, like
images or tables alongside text.

As part of an initial study, the decision was made to focus on
pre-retrieval, at-retrieval, post-retrieval, and generation, however
future work will dive deeper into identified methods in advanced
augmentation. We describe our implementation of each of the leveraged
methods here:

\subsection{Pre-retrieval}

Pre-retrieval techniques play a crucial role in enhancing the
effectiveness of information retrieval systems by expanding and refining
queries prior to the standard retrieval process used for final
generation. In this context, query expansion serves as a powerful tool
to improve the quality and relevance of retrieved information. We
explore two distinct approaches to query expansion: upward abstraction
and downward abstraction.

\begin{enumerate}
\def\labelenumi{\arabic{enumi}.}
\item
  \textbf{Query Expansion -- Upward Abstraction}: To investigate the
  impact of abstraction in query processing, we implement a method based
  on Zheng et al.'s Step-Back method {[}17{]}. Step-Back is an automated
  prompting technique designed to assist LLMs with the process of
  abstracting higher-level concepts before using reasoning to answer
  questions. This method follows a two-step process. In the first step,
  upon receiving a question from the user, the LLM is prompted to
  explore higher-level principles associated with the question by
  generating what is called a Step-Back question. For example, if a user
  asks, ``how much paint do I need to paint this wall?'' Step-Back might
  generate ``what methods are used for calculating square footage of a
  surface?'' Both the Step-Back question and the original question are
  then used for retrieval of relevant documents. In the reasoning step,
  the LLM generates an answer to the original question using documents
  retrieved for both the Step-Back and original question as context.
\item
  \textbf{Query Expansion -- Downward Abstraction:} To explore how
  abstraction performs when applied in the opposite direction, by
  decomposing a question, we adapt the method proposed in Kim et al.'s
  Tree of Clarifications paper {[}18{]}. This technique is a framework
  which, when prompted by a user's question, creates a tree data
  structure where the root node is the original question, and the
  children are question-answer pairs, each consisting of a disambiguated
  question decomposed from the original by the LLM, and its
  corresponding answer. For example, one question from our dataset asks,
  ``Did American Airlines Group, Inc. issue any new debt instruments?''
  This original question serves as the root node of the tree, while the
  LLM generates question-answer pairs as child nodes: \{``Has American
  Airlines Group, Inc. modified any existing debt covenants?'': ``Yes,
  American Airlines Group, Inc. has recently renegotiated debt terms''\}
  and \{``Are there any upcoming changes to the maturity schedule of
  American Airlines' long-term debt?'': ``American Airlines, through the
  2013 Credit Agreement, extended the maturity of \$1.0 billion in term
  loans under the 2013 Term Loan Facility''\}. The various nodes of the
  tree are combined to determine a comprehensive answer to the original
  question. Unlike the original, our implementation uses a fixed passage
  for one-shot learning and does not prune branches of the generated
  tree.
\end{enumerate}

\subsection{At-retrieval}

At-retrieval techniques are pivotal in refining the process of
information retrieval by optimizing how queries interact with the
retrieval system. This section focuses on evaluating retrieval types and
fine-tuning embedding models to enhance the precision and relevance of
retrieved information.

\begin{enumerate}
\def\labelenumi{\arabic{enumi}.}
\item
  \textbf{Retrieval Type:} To compare the effectiveness of different
  retrieval approaches, we conducted experiments using both dense and
  sparse retrieval methods. For dense retrieval, we employed a
  bi-encoder architecture where both queries and documents were embedded
  into a shared vector space using an embedding model. Sparse retrieval,
  on the other hand, was implemented using BM25, a traditional
  term-based retrieval system which ranks documents based on lexical
  overlap with the query.
\item
  \textbf{Fine-Tuning -- Embedding Model:} To understand the effects of
  better-quality embeddings, we fine-tune the bge-small-en embedding
  model {[}27{]} by adjusting the top 8 layers using a contrastive loss
  objective. This approach involves creating 149 distractor passages for
  each positive passage, ensuring the model learns to distinguish
  relevant documents from irrelevant ones effectively. We employ the
  AdamW optimizer with a learning rate of 1e-6, a batch size of 4, and a
  weight decay of 0.01 to improve generalization. Fine-tuning was
  conducted over 3 epochs, balancing training duration with the risk of
  overfitting. This process enhanced the model's ability to generate
  task-specific embeddings, improving retrieval quality in downstream
  evaluations.
\end{enumerate}

\subsection{Post-retrieval}

Post-retrieval techniques enhance the quality of responses by refining
the retrieved information to capture additional relevant information. We
explore iterative retrieval methods, both fixed and adaptive, to improve
accuracy and completeness. Fixed iteration uses a set number of cycles
for consistent refinement, while adaptive iteration adjusts dynamically
based on evaluations of completeness. These approaches aim to optimize
generation, ensuring the correct information is retrieved.

\begin{enumerate}
\def\labelenumi{\arabic{enumi}.}
\item
  \textbf{Iterative Retrieval - Fixed:} To explore impacts of
  post-retrieval iteration, we deploy a method mirroring Shao et al.'s
  ITER-RETGEN {[}19{]}. The core idea of this method is to iteratively
  refine the generated response by repeatedly retrieving additional
  information based on previously generated answers. Initially, the
  process resembles a traditional RAG pipeline: information relevant to
  the original prompt is retrieved, and a preliminary answer is
  generated from this retrieved content. In subsequent iterations, this
  preliminary answer---combined with the original prompt---is fed back
  into the retriever, allowing it to identify and retrieve additional
  content that was previously missed due to limitations in initial
  embedding representations. Each retrieved set of new information is
  then used to generate an updated answer. ITER-RETGEN is unique in that
  it repeats this process a fixed number of times, established during
  implementation. Our implementation iterates on this process 3 times,
  after which the final generation is taken as the answer.
\item
  \textbf{Iterative Retrieval - Adaptive:} Here we deploy a technique to
  contrast static iteration with dynamic iteration of the retrieval
  step, heavily inspired by Khattab et al.'s Demonstrate-Search-Predict
  {[}20{]}. Unlike the fixed method, which performs a predetermined
  number of iterations, this approach dynamically adjusts retrieval
  iterations based on intermediate evaluations of completeness and
  confidence. Initially, demonstrations are generated using a small set
  of predetermined, hard-coded example queries. These demonstrations
  provide guidance on how to handle similar input queries, informing the
  decomposition and retrieval strategy. Following these examples, the
  system decomposes the input query into logical sub-questions and
  iteratively retrieves relevant information for each sub-question. At
  each retrieval iteration, the system evaluates whether the retrieved
  information sufficiently addresses the sub-questions, dynamically
  deciding whether additional retrieval iterations are necessary. Once
  sufficient information has been gathered, the retrieved documents,
  sub-questions, original query, and additional context are combined
  into a comprehensive prompt, from which the final answer is generated.
\end{enumerate}

\subsection{Generation}

Evaluating fine-tuning of the generation model aims to improve the
accuracy and relevance of its outputs. By adjusting model parameters, we
seek to enhance its ability to produce contextually precise responses.

\begin{enumerate}[label=\phantom{\arabic*.},widest=2,nosep]
\item \textbf{Fine-Tuning -- Generation Model:} To understand the effects of
parametric adjustment of an LLM towards our objective, we fine-tune
LLaMA 3.2 Instruct 3B {[}24{]} by adjusting the top 2 layers and the
model head using a custom loss function. This loss combines
cross-entropy with a weighted soft n-gram match, balancing 75\%
cross-entropy and 25\% n-gram similarity to encourage more contextually
accurate generations. We use the AdamW optimizer with a learning rate of
1e-6, applying cosine warmup to stabilize early training. Fine-tuning is
performed using Distributed Data Parallel (DDP) across 2 NVIDIA A40
GPUs, with a batch size of 1 per GPU and 16-step gradient accumulation,
resulting in an effective batch size of 32. Weight decay is set to 0.01,
and training is conducted for 4 epochs, ensuring learning without
overfitting.
\end{enumerate}

\section{Data}

Data engineering efforts for Q\&A RAG systems typically involve the
curation of two types of data: a set of questions for training and
evaluation and a corpus to retrieve from and reason against when
deriving answers {[}23{]}. To generate a set of questions for our use
case we compile a targeted list of risk indicators used for prior supply
chain risk assessments {[}28, 29, 30{]}. We then decompose those
indicators into a series of questions which are typically researched and
answered by analysts to qualify the vulnerability level of a supply
chain. These questions cover a broad range of potential vulnerabilities
involving topics such as critical mineral usage, geographic
concentration, research and development investment, child labor
exposure, and other topics related to the operations of a business and
the health of their supply chains. For our corpus, we elect to use SEC
filings because they are publicly available, offer valuable insights
into supply chain risks, and are abundant enough to facilitate effective
tuning of the models in our system.

To generate the details necessary to allow for experimentation with
fine-tuning, we extend the set of questions to triplets containing the
original question, a supporting passage from the corpus which contains
the answer (either explicit or derived), and a correct answer to the
question {[}14{]}. We accomplish this through two methods:

\begin{enumerate}
\def\labelenumi{\arabic{enumi}.}
\item
  \textbf{Human Annotation:} A group of subject matter experts (SMEs)
  are provided access to the SEC's Electronic Data Gathering, Analysis,
  and Retrieval (EDGAR) database {[}21{]} and asked to document answers
  to the set of questions provided. EDGAR is the SEC's repository for
  financial disclosures and the SMEs are asked to only use these
  documents as their source of information for the requested task. Along
  with the determined answers, SMEs are also asked to record what they
  feel to be the best passage out of all the filings reviewed which best
  assist in answering the question. From this exercise, SMEs identified
  that four types of disclosures were required to answer the questions
  specific to our use case -- 10-Ks, 10-Qs, 8-Ks, and 14-As.
\item
  \textbf{Synthetic Generation:} Due to the volume limitations of
  creating this data manually and the need for large datasets for
  fine-tuning {[}15{]}, we deploy synthetic methods to scale our data
  creation beyond human annotation. Specifically, we leverage the method
  proposed by Saad-Falcon, Khattab, Potts, and Zaharia {[}22{]} which
  uses a large parameter LLM to generate question, passage, and answer
  triplets from an in-domain corpus. For our purposes we leverage GPT-4o
  and a collection of the types of disclosures identified to be relevant
  through human annotation.
\end{enumerate}

The integration of these methods provides a comprehensive and diverse
dataset, instrumental to tuning and testing the proposed system's
retrieval and generation capabilities.

\section{Experiments}

The following experimental design is constructed to empirically measure
the time required for an LLM to ingest new supply chain information
under performance thresholds for effectively identifying potential risks
and vulnerabilities. Accomplishing this begins with a baseline RAG
implementation to act as a basis for this measurement. To allow for
controlled and rapid experimentation we deploy a mechanism to simulate
an event of interest occurring which results in new information flowing
to our system. Finally, we extend the deployed RAG instance to support a
modular design that allows for the techniques described above to be
leveraged in various combinations and study their effects.

The table below provides a detailed overview of this experimental setup.
Each row corresponds to a specific RAG technique, including dense
retrieval (alpha), fine-tuning the retriever (FTR) and generator (FTG),
upward/downward abstraction (UA/DA), and fixed/adaptive iterative
retrieval (FIR/AIR). A \textquotesingle1\textquotesingle{} in a row
indicates that the corresponding technique was utilized in a particular
configuration. The columns, labeled C1 to C30, represent unique
configurations of these techniques, illustrating how different
combinations are tested to assess their impact on the
system\textquotesingle s performance.

\begin{table}[htbp]
  \caption{Experimental Configurations}
  \label{tab:configs}
  \centering
  \vspace{6pt} 
  \includegraphics[width=\linewidth]{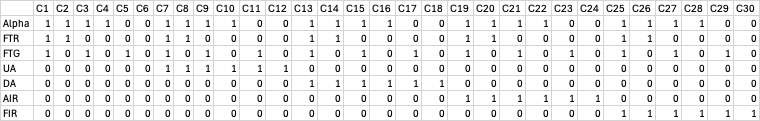}
\end{table}

We leverage LlamaIndex {[}26{]} to facilitate the ingestion, indexing,
and retrieval of supply chain data, enabling a scalable and modular data
pipeline. Weaviate {[}25{]} was chosen as the vector store for its
open-source nature and support for hybrid retrieval, combining semantic
and lexical search. bge-small-en was selected as the embedding model for
its balance of computational efficiency and retrieval effectiveness. For
generation, we use Llama 3.2 3B Instruct, as instruct models are
well-suited for robust RAG prompt tuning {[}16{]}. A smaller model was
chosen mainly due to hardware limitations but also to reduce the
likelihood of task-specific pretraining biases and to better demonstrate
the impact of our augmentation methods.

To simulate the real-time ingestion of new, relevant events into an LLM,
we leverage the SEC's EDGAR system. Using custom scripting, we automate
the retrieval of filings for a predetermined set of companies that align
with our evaluation questions. These filings are stored in a MongoDB
instance along with relevant metadata, and a subset is encoded and
ingested into the vector store as an initial knowledge base. During each
experimental run, we systematically ingest the remaining disclosures
into the vector store and measure the system's ability to integrate and
adapt to the newly introduced information. This setup allows us to
assess the efficiency and effectiveness of updating an LLM's knowledge
in response to real-world outcomes -- in this case, a company filing a
corporate disclosure which may contain supply chain implications.

To enable experimentation with the methods introduced in Section 4, we
extend our baseline RAG architecture to incorporate modular insertion
points that allow for targeted augmentations at different stages of the
pipeline. These insertion points are strategically implemented at key
phases, including query encoding, pre-retrieval, retrieval,
post-retrieval, and generation. This modular design enables different
augmentation techniques to be implemented as interchangeable modules,
which can be toggled on or off at various intervals to assess their
impact, both in isolation and in combination. For the purposes of this
study, a specific set of deployed techniques is referred to as a
configuration, with each experimental run deploying a distinct
configuration for evaluation. Our experiments methodically analyze the
effects of 30 such configurations, which are detailed in Table 1.

\subsection{Metrics}

Formal metrics are also provided here to further define the presented
objective as well as instantiate a definition of success and how it will
be measured. To measure the effectiveness of retrieval
(\(p_{\eta}(z \mid x)\)), we deploy a combination of Normalized
Discounted Cumulative Gain (nDCG) {[}9{]}, Hit Rate, Average Rank, and
Mean Reciprocal Rank (MRR) {[}10{]}. nDCG is a ranking quality metric
that discounts the relevance of items based on their positions in the
list and then normalizes the total gain to a perfect ordering. Hit Rate
is an accuracy metric used to determine whether or not the best
supporting passage was included within the retrieved information.
Average Rank assists in identifying the quality of retrieved ordering by
measuring the mean of the rank of relevant items across all queries. MRR
is the average of the reciprocal of the ranks at which the first
relevant item is found for each query.

To assess the quality of generations in our experiments (\(M)\) we
evaluate the outputs using Recall-Oriented Understudy for Gisting
Evaluation (ROUGE) {[}11{]}, BiLingual Evaluation Understudy (BLEU)
{[}12{]}, Exact Match, and Semantic Similarity {[}13{]}. ROUGE and BLEU
are n-gram matching metrics which evaluate the proximity of a predicted
answer from an LLM to the correct labeled data. An important distinction
between these measures is BLEU penalizes for n-gram ordering whereas
ROUGE does not. As its name implies, Exact Match is a count of generated
answers that exactly match the labeled data. This is often the strictest
of metrics in a study like ours, particularly when lengthy responses
from the LLM are expected. To account for answers which may have
different wording (n-grams) but maintain the correct semantic meaning as
the answer, we measure Semantic Similarity which assesses how closely
two pieces of text convey the same meaning based on their contextual and
conceptual alignment rather than exact word overlap. This is achieved by
representing text as high-dimensional vectors and measuring their
distance in embedding space, where semantically similar texts have
smaller distances.

In an effort to not let these issues skew the metrics above, we also
measure a count of instances where the model either returns a blank
response or an indication that it cannot answer the question with its
existing or provided knowledge. Our final metric is time which is
measured as the number of seconds required to generate answers to our
entire evaluation dataset. This includes the time necessary to
facilitate all processing in our pipeline including fine-tuning, where
applied.

\begin{figure}[htbp]
\centering
\includegraphics[width=\linewidth]{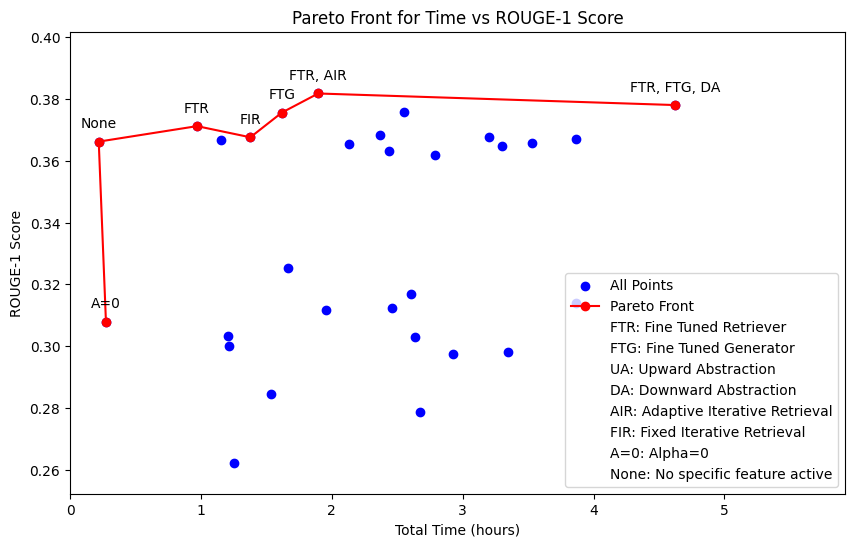}
\caption{Trade Space Evaluation - ROUGE-1 and Time in hours}
\end{figure}

\section{Results}

The results of the study are explored in the following two subsections.
In the first subsection we examine the trade space specific to the dual
objective of timeliness and quality described in Section 2. In the
second subsection, we delve into key insights regarding the impacts and
effects of certain combinations of techniques, including retrieval type
where alpha = 1 is dense retrieval, fine-tuning, query expansion, and
iterative retrieval, as detailed in Section 4. It should be noted that
all measurements of the techniques mentioned in 6.2 below are averages
of the performance of all configurations which contain that technique.
The one exception to this is the ``None'' category in Figures 4c and 4d
which is the average of the unaltered baseline model, both with and
without dense retrieval. We do this to demonstrate the impacts of the
highlighted techniques compared to a RAG implementation which uses
frozen pretrained models. We conclude this section with a qualitative
measurement of the results of our experiments along with some key
finding by SMEs.

\subsection{Trade Space Analysis}

Our analysis identifies the 7 configurations that form the optimal
boundary of our trade space when evaluating dual objectives: quality and
timeliness. Here, we leverage ROUGE-1 as the quality metric and time in
hours as the timeliness variable. ROUGE-1 is chosen here due to the
importance of key terms, such as product or company names, which must be
preserved in the output. As mentioned previously, time in this context
encompasses the full pipeline for generating answers to the evaluation
questions, including all processing and fine-tuning. Our best-performing
configuration, identified as the highest point on the y-axis of Figure
2, consists of fine-tuning the retriever combined with adaptive
iterative retrieval. This combination likely excels because fine-tuning
enhances initial retrieval accuracy, and adaptive retrieval offers the
flexibility to iterate as many times as necessary to gather the required
information. Unsurprisingly, the unaltered configuration (no fine-tuning
or retrieval techniques applied) scores highest for timeliness,
represented by the leftmost point on the x-axis of the same Figure.
Other points along the optimal boundary demonstrate the trade space
considerations when deploying a RAG system, allowing for improvements
based on implementation priorities. For example, fine-tuning the
retriever offers a good tradeoff between timeliness and accuracy by
enhancing model performance without significantly increasing latency.
When timeliness is paramount, one should move left and down along the
boundary line, while prioritizing quality requires shifting right and
up. Points falling below and to the right of the identified exemplars
highlight diminishing returns, with the rightmost configuration
sacrificing quality for a disproportionately large increase in
processing time, while the bottom-most configuration exhibits a sharp
drop in quality without any corresponding improvement in timeliness.

\subsection{Impact of Techniques on Performance}

\subsubsection{Recovery from Poor Retrieval through Good Generation}

Our results highlight the importance of good retrieval and demonstrate
that recovering from poor performance in this area can be quite
challenging. Specifically, as shown in Figure 3, no combination of
tuning strategies deployed can beat the performance gains of fine-tuning
the retriever. Interestingly, we find that when combining fine-tuning of
the retriever with fine-tuning of the generator, these gains are
diminished slightly, on average for the ROGUE-1 metric and Semantic
Similarity metric. However, we see a slight overall gain with respect to
BLEU when combining these two techniques.

\subsubsection{Recovery from Bad Generation through Good Retrieval}

As illustrated in Figure 3, consistent performance gains can be observed
over the base model from fine-tuning the generator. In each of these
cases however we find that these gains can be recouped, and in fact
exceeded in some cases, by fine-tuning the retriever, either in lieu of
or combination with fine-tuning the generator. The process of
fine-tuning the generator also typically takes considerably more time
than the retriever. This suggests that in a resource constrained
environment, fine-tuning of a retriever should be prioritized.

\begin{figure}[htbp]
\centering
\includegraphics[width=\linewidth]{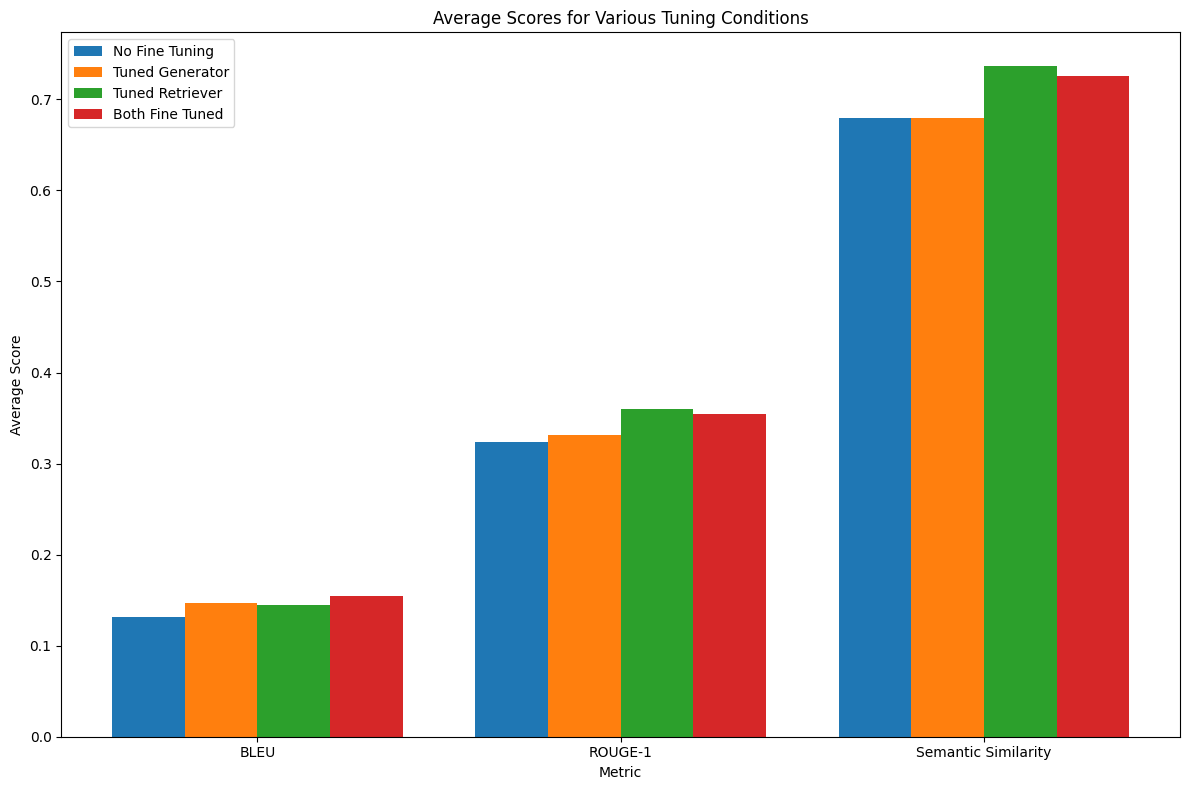}
\caption{Impacts of Fine-Tuning}
\end{figure}

\subsubsection{Recovery through Optimization (RAG Techniques)}

Figure 4a, shows that none of the tested techniques outperform
fine-tuning the retriever, on average. This once again highlights the
importance of good retrieval and demonstrates that you cannot fully
recover these gains using alternative methods, at least the ones tested
in this study. The closest candidates for recovery, however, are
iterative retrieval techniques which demonstrate the potential for
partial recovery. Alternatively, as depicted in Figure 4b, these same
techniques either match or slightly underperform the performance of
fine-tuning the generator. This suggests that you can indeed recover the
performance gains from not fine-tuning the generator through iterative
retrieval. This is a significant finding as it suggests these techniques
as an alternative to fine-tuning the generator which is costly in both
time and computation.

\subsubsection{Query Expansion: Abstracting Upward vs. Downward}

As shown in Figure 4c, upward abstraction considerably underperforms in
comparison to downward abstraction. In fact, configurations using upward
abstraction are consistently the lowest performing, on average,
including that of the unaltered base models. This is likely due to the
method introducing higher level details to prompts which do not directly
help in answering the underlying questions. Downward abstraction does
not introduce noise in this way because the additional details generated
are directly decomposed from the original question. As a result,
downward abstraction proves to be a more effective query expansion
method for our use case.

\subsubsection{Iterative Retrieval: Fixed vs. Adaptive}

As illustrated in Figure 4d, adaptive iterative retrieval consistently
outperforms fixed iterative retrieval, though the difference between the
two remains relatively close. In addition to this improvement, it also
demonstrates a clear advantage over the unaltered base models, further
highlighting its role in enhancing retrieval performance. This suggests
that adaptively determining the number of retrieval steps based on
context leads to more effective information retrieval than either a
fixed iterative approach or relying solely on an unaltered RAG
implementation. This validates the influence of adaptive iterative
retrieval on the highest-performing configuration, demonstrating that
its strong individual performance is a key factor contributing to the
overall effectiveness of the combined approach (see Section 6.1).

\begin{figure}[htbp]
\centering
\includegraphics[width=\linewidth]{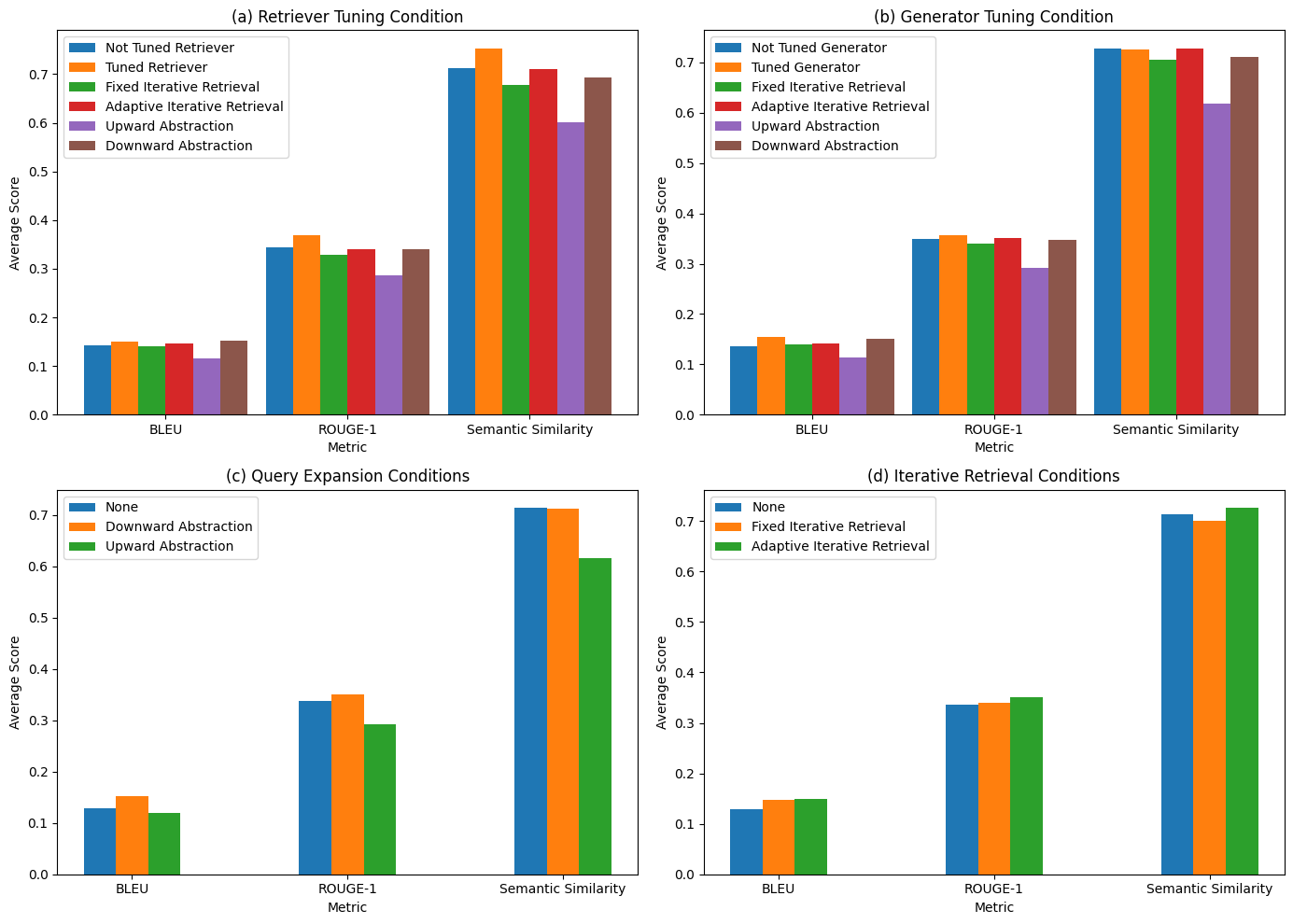}
\caption{Impacts of RAG Techniques}
\end{figure}

\subsection{Human Evaluations}

In addition to assessing performance using quantitative metrics such as
ROUGE, BLEU and Semantic Similarity, we compare the outputs of our
top-performing configurations, downward abstraction with fine-tuned
retriever and adaptive retrieval with fine-tuned retriever,
qualitatively through a team of subject matter experts. To achieve this,
experts are presented with the evaluated questions, retrieved passages,
and generated answers from each of these configurations, along with
those from the unaltered base model, and asked to qualitatively assess
their outputs. The questions are designed to evaluate a firm's supply
chain risks based on information from their recent SEC filings, and the
experts are economists and industry analysts who work on supply chain
risk assessments. Specifically, the experts are asked to document if, in
their expert judgement, they feel the generated answer is correct and to
flag if there are instances where they feel the system-generated answer
is in fact more accurate than the answer provided through our
human-annotated dataset, suggesting the model is correct over the human.
These experts are also asked to indicate if they feel each question is
particularly complex or important.

\begin{figure}[H]
  \centering
  \begin{minipage}{0.48\linewidth}\centering
    \includegraphics[width=\linewidth]{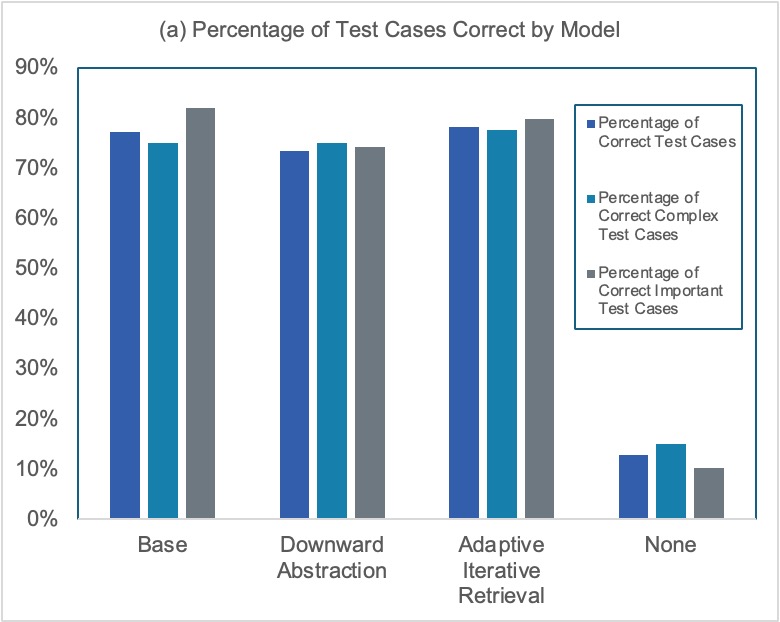}
  \end{minipage}\hfill
  \begin{minipage}{0.48\linewidth}\centering
    \includegraphics[width=\linewidth]{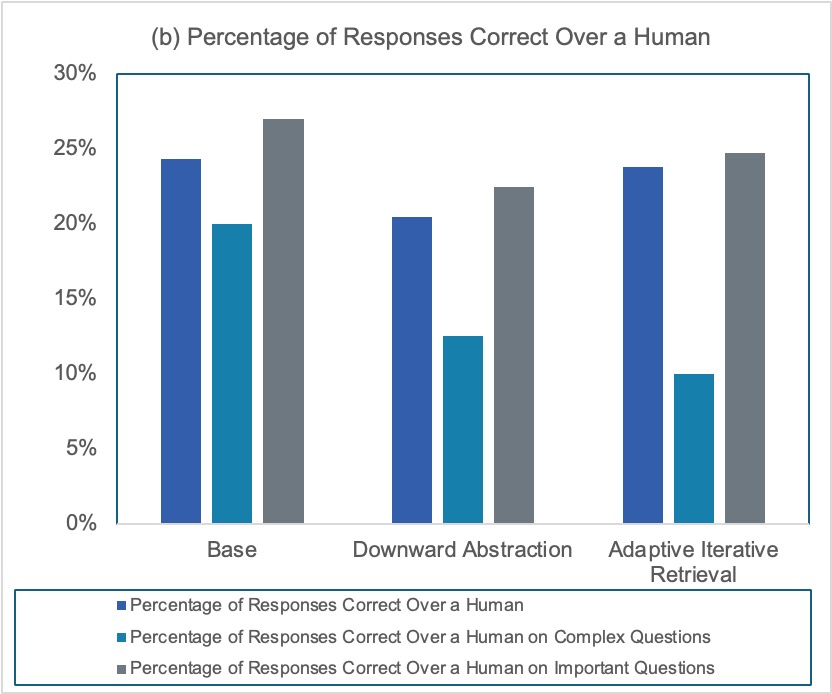}
  \end{minipage}
  \caption{Qualitative Assessment}\label{fig:qualitative}
\end{figure}

As shown in Figure 5, among the three configurations evaluated, adaptive
iterative retrieval produces the most correct answers, achieving 78\%
accuracy across 210 entries. We use four human supply chain experts in
our testing. It also performs best on complex questions, likely due to
its ability to simulate reasoning through multiple rounds of retrieving
and assessing relevant information. Interestingly, the base model
performs slightly better on questions labeled as important, suggesting
that such questions may sometimes be more straightforward (less
complex). This finding underscores the need for retrieval-based systems
to balance deeper exploration for complex queries with more direct
retrieval for simpler but high-priority questions.

Our analysis reveals that only 13\% of the evaluated questions
demonstrate cases where human-generated answers are superior to all
three tested configurations. Notably, as illustrated in Figure 5, the
base model outperforms human responses in 24\% of all cases---the
highest among the configurations. The weakest configuration, in
contrast, achieves a 10\% correctness rate over human responses
specifically in high-complexity questions. While further error analysis
is needed for definitive conclusions, these findings suggest that RAG
systems have the potential not only to reduce the time required for this
type of analysis but also to enhance accuracy compared to traditional
manual efforts.

We find that the tested RAG configurations excel at locating the
proverbial ``needle in a haystack,'' often unearthing relevant
information in unexpected places. When recent filings lack the needed
data, the system frequently searches for older filings rather than
giving up. In some cases, historical information proves essential---such
as exposing a firm that disguised its People's Republic of China (PRC)
origins by replacing its China-based board with a U.S.-based
directorate. Relying solely on the latest filing would have produced a
sanitized version of that firm's history. At the same time, the system
sometimes over-relies on outdated filings. For example, it struggles to
identify competitors to a prominent electric vehicle (EV) company,
largely because that firm's SEC filings at a certain point in time
stopped naming competitors. Other EV manufacturers, however, do identify
competing brands in their annual reports. By fixating on the older
filings instead of consulting those more recent disclosures, the system
disproportionately emphasizes European and Japanese luxury
marques---overlooking newer entrants in the global electric and
autonomous vehicle space, including both U.S. and Chinese technology and
auto manufacturers.

While the tested RAG configurations generally excel at finding relevant
content in older documents, they struggle with comparing year-over-year
filings, typically detecting increases or decreases only when ``growth''
is explicitly mentioned. They also identify risk-related language but do
not assess the actual level of exposure. Across industries, natural
disasters, supply-chain disruptions, competition, and government
regulation feature prominently in filings---yet the system cannot
distinguish routine boilerplate from true existential threats without
further context. Visibility also matters, as the system consistently
flags well-known firms that disclose detailed labor or human rights
practices, while smaller or less-scrutinized firms may slip under the
radar. In one notable instance, the system erred on a question of
foreign ownership by assuming a CEO's nationality based on his name,
rather than recognizing an actual foreign sovereign wealth fund's
investment. Such errors underscore the need for future work to refine
how the system evaluates and interprets these documents.

\section{Conclusion}

This research demonstrates the potential of RAG systems, coupled with
web-scraping technologies, to enhance real-time supply chain analysis.
This paper evaluates the interplay between retrieval and generation
techniques. The key findings highlight that fine-tuning the retriever
consistently yields the greatest performance gains in accurately
identifying supply chain disruptors. While iterative retrieval
techniques offer partial recovery from suboptimal configurations, our
analysis confirms that high-quality retrieval is essential for
maximizing both accuracy and efficiency. Through human assessment we
show the RAG system's ability to surface critical supply chain insights,
often identifying relevant information overlooked in manual reviews.

Given the increasing demand for more timely and detailed supply chain
intelligence---especially within the context of U.S. economic and
national security policies---our approach presents a scalable and
automated solution for aggregating regulatory filings into actionable
insights. This has significant implications for policymakers,
researchers, and industry professionals who rely on rapid and accurate
assessments of global trade dynamics.

Building on these findings, future research will focus on optimizing the
system's ability to extract and relate firm- and product-specific
information while also exploring the temporal weighting of sources. We
anticipate that integrating knowledge graphs will strengthen the
system's capacity to understand and contextualize relationships between
firms, suppliers, and geopolitical factors. Additionally, addressing the
issue of overreliance on older documents could be achieved through
temporal weighting of sources or reranking with this intent. Expanding
real-time data ingestion through internet augmentation will further
reduce latency and improve responsiveness to emerging disruptions.

Furthermore, we aim to examine how RAG techniques perform relative to
recent innovations in pretraining, clarifying the conditions under which
these techniques provide the most value. Due to resource constraints and
the extensive processing time required for each experimental run, we
were unable to thoroughly investigate the statistical significance of
our findings---a limitation we plan to address in future research. Such
efforts will strengthen the validity of our conclusions and yield deeper
insights into the comparative performance of RAG systems. Ultimately,
refining these techniques will enhance the effectiveness of LLMs in
improving supply chain visibility and resilience within an increasingly
complex global landscape.

\section*{Acknowledgements}

This work was funded by the MITRE Independent Research and Development
Program. Approved for Public Release; Distribution Unlimited. PRS
Release Number: 25-0864.\textbf{\hfill\break
}

\section*{References}

{[}1{]} ``America First Trade Policy.'' The White House, January 20,
2025.
https://www.whitehouse.gov/presidential-actions/2025/01/america-first-trade-policy/

{[}2{]} ``Unleashing American Energy.'' The White House, January 20,
2025.
https://www.whitehouse.gov/presidential-actions/2025/01/unleashing-american-energy/

{[}3{]} U.S. Department of Homeland Security, ``Uyghur Forced Labor
Prevention Act: Frequently Asked Questions,'' \emph{Department of
Homeland Security}, Mar. 2023. {[}Online{]}. Available:
\url{https://www.dhs.gov/uflpa-frequently-asked-questions}

{[}4{]} R. Baldwin and R. Freeman, ``Risks and global supply chains:
What we know and what we need to know,'' Annual Review of Economics,
vol. 14, no. 1, pp. 153--180, 2022. {[}Online{]}. Available:
https://doi.org/10.1146/annurev-economics-051420-113737

{[}5{]} Purdue University, ``GTAP 11 Database,'' \emph{Global Trade
Analysis Project}, 2022. {[}Online{]}. Available:
\url{https://www.gtap.agecon.purdue.edu/databases/v11/}

{[}6{]} U.S. Bureau of Economic Analysis, "Input-Output Accounts Data,"
\emph{U.S. Department of Commerce}, 2023. {[}Online{]}. Available:
\url{https://www.bea.gov/industry/input-output-accounts-data}

{[}7{]} P. Antràs and D. Chor, "Global Value Chains," NBER Working Paper
No. 28549, Mar. 2021. {[}Online{]}. Available:
https://www.nber.org/papers/w28549

{[}8{]} P. Lewis et al., "Retrieval-Augmented Generation for
Knowledge-Intensive NLP Tasks," in \emph{Advances in Neural Information
Processing Systems (NeurIPS), 2020}, pp. 6459-6475

{[}9{]} K. Järvelin and J. Kekäläinen, ``Cumulated gain-based evaluation
of IR techniques,'' \emph{ACM Trans. Inf. Syst.}, vol. 20, no. 4, pp.
422--446, Oct. 2002. {[}Online{]}. Available:
\url{https://doi.org/10.1145/582415.582418}.

{[}10{]} E. M. Voorhees, ``The TREC-8 question answering track report,''
in \emph{Proc. 8th Text Retrieval Conf. (TREC-8)}, Gaithersburg, MD,
USA, 1999, pp. 77--82.

{[}11{]} C. Lin, "ROUGE: A Package for Automatic Evaluation of
Summaries," \emph{Text Summarization Branches Out}, Barcelona, Spain,
2004, pp. 74--81.

{[}12{]} K. Papineni, S. Roukos, T. Ward, and W. Zhu, "BLEU: a Method
for Automatic Evaluation of Machine Translation," \emph{Proceedings of
the 40th Annual Meeting of the Association for Computational
Linguistics}, Philadelphia, PA, USA, 2002, pp. 311--318.

{[}13{]} P. Resnik, "Using information content to evaluate semantic
similarity in a taxonomy," in \emph{Proceedings of the 14th
International Joint Conference on Artificial Intelligence (IJCAI)},
Montreal, Quebec, Canada, 1995, pp. 448--453. {[}Online{]}. Available:
\url{https://arxiv.org/abs/cmp-lg/9511007}

{[}14{]} H. Du, H. Zhang, and D. Zhao, "Evidence-Enhanced Triplet
Generation Framework for Hallucination Alleviation in Generative
Question Answering," arXiv preprint arXiv:2408.15037, Aug. 2024

{[}15{]} I. Vieira, W. Allred, S. Lankford, S. Castilho, and A. Way,
"How Much Data is Enough Data? Fine-Tuning Large Language Models for
In-House Translation: Performance Evaluation Across Multiple Dataset
Sizes," in \emph{Proceedings of the 16th Conference of the Association
for Machine Translation in the Americas (Volume 1: Research Track}),
Chicago, USA, 2024, pp. 236--249.

{[}16{]} Z. Wei, W.-L. Chen, and Y. Meng, "InstructRAG: Instructing
Retrieval-Augmented Generation via Self-Synthesized Rationales," in
\emph{Proceedings of the 13th International Conference on Learning
Representations, 2025}.

{[}17{]} H. S. Zheng et al., ``Take a step back: Evoking reasoning via
abstraction in large language models,'' arXiv, Oct. 2023. {[}Online{]}.
Available: \url{https://arxiv.org/abs/2310.06117}.

{[}18{]} G. Kim, S. Kim, B. Jeon, J. Park, and J. Kang, ``Tree of
Clarifications: Answering Ambiguous Questions with Retrieval-Augmented
Large Language Models,''~arXiv.org, 2023. {[}Online{]}.
https://arxiv.org/abs/2310.14696.

{[}19{]} Z. Shao, Y. Gong, Y. Shen, M. Huang, N. Duan, and W. Chen,
``Enhancing retrieval-augmented large language models with iterative
retrieval-generation synergy,'' arXiv, vol. abs/2305.15294, 2023.
{[}Online{]}. Available: \url{https://arxiv.org/abs/2305.15294}.

{[}20{]} O. Khattab, K. Santhanam, X. L. Li, D. L. W. Hall, P. Liang, C.
Potts, and M. A. Zaharia, ``Demonstrate-search-predict: Composing
retrieval and language models for knowledge-intensive NLP,''
\emph{arXiv}, vol. abs/2212.14024, 2022. {[}Online{]}. Available:
\url{https://arxiv.org/abs/2212.14024}.

{[}21{]} U.S. Securities and Exchange Commission, "Electronic Data
Gathering, Analysis, and Retrieval (EDGAR) database." {[}Online{]}.
Available: \url{https://www.sec.gov/edgar.shtml}. Accessed: Sep 2024.

{[}22{]} J. Saad-Falcon, O. Khattab, C. Potts, and M. Zaharia, ``Ares:
An automated evaluation framework for retrieval-augmented generation
systems,'' in \emph{Proc. 2024 Conf. North Amer. Chapter Assoc. Comput.
Linguist.: Human Lang. Technol. (NAACL-HLT)}, Mar. 2024. {[}Online{]}.
Available: \url{https://arxiv.org/abs/2311.09476v2}.

{[}23{]} J. Lala, O. O'Donoghue, A. Shtedritski, S. Cox, and S. G.
Rodriques, ``PaperQA: Retrieval-Augmented Generative Agent for
Scientific Research,'' in \emph{Proc. 23rd Int. Conf. Model. Appl.
Simul.}, Dec. 2023, pp. 1--20. doi:
\href{https://doi.org/10.46354/i3m.2024.mas.021}{10.46354/i3m.2024.mas.021}.

{[}24{]} Meta, \emph{Llama 3.2 3B Instruct}, Hugging Face, Sep. 25,
2024. {[}Online{]}. Available:
\url{https://huggingface.co/meta-llama/Llama-3.2-3B-Instruct}. Accessed:
Oct 2024.

{[}25{]} E. Dilocker, B. van Luijt, B. Voorbach, M. S. Hasan, A.
Rodriguez, D. A. Kulawiak, M. Antas, and P. Duckworth, \emph{Weaviate},
GitHub repository, 2023. {[}Online{]}. Available:
\url{https://github.com/weaviate/weaviate}. Accessed: Sep 2024.

{[}26{]} J. Liu, \emph{LlamaIndex}, Nov. 2022. {[}Online{]}. Available:
\url{https://github.com/jerryjliu/llama_index}. doi:
10.5281/zenodo.1234.

{[}27{]} S. Xiao, Z. Liu, P. Zhang, and N. Muennighoff, "BGE: BAAI
general embedding," \emph{GitHub repository}, 2023. {[}Online{]}.
Available: \url{https://github.com/FlagOpen/FlagEmbedding}. Accessed:
Oct. 2024.

{[}28{]} National Economic Council and National Security Council,
2021--2024 Quadrennial Supply Chain Review, The White House, Washington,
DC, 2024. {[}Online{]}. Available:
\url{https://bidenwhitehouse.archives.gov/wp-content/uploads/2024/12/20212024-Quadrennial-Supply-Chain-Review.pdf}.

{[}29{]} R. A. Martin, \emph{Trusting our supply chains: A comprehensive
data-driven approach}, The MITRE Corporation, 2020. {[}Online{]}.
Available:
\url{https://www.mitre.org/sites/default/files/2021-11/prs-20-01465-37-trusting-our-supply-chains-comprehensive-data-driven-approach.pdf}.

{[}30{]} N. Bashain, \emph{DoD advanced battery supply chain risk
assessment, Mil. Power Sources Consortium Submission}, 2023.
{[}Online{]}. Available:
\url{https://apps.dtic.mil/sti/trecms/pdf/AD1214085.pdf}.

{[}31{]} D. Edge et al., "From Local to Global: A Graph RAG Approach to
Query-Focused Summarization," arXiv preprint arXiv:2404.16130, Apr.
2024. {[}Online{]}. Available:
https://arxiv.org/abs/2404.16130\hspace{0pt}

\end{document}